\documentclass[journal=jacsat,manuscript=article]{achemso}

\usepackage[version=3]{mhchem} 
\usepackage{physics}
\usepackage{booktabs}
\usepackage[percent]{overpic}
\usepackage{subcaption}
\usepackage{hyperref}
\usepackage{color}
\usepackage{bm}
\usepackage[utf8]{inputenc}

\newcommand{\bh}{\mbox{\protect\boldmath $h$}}

\newcommand{\bK}{\mbox{\protect\boldmath $K$}}

\newcommand{\bR}{\mbox{\protect\boldmath $R$}}

\newcommand{\bA}{\mbox{\protect\boldmath $A$}}

\newcommand{\balpha}{\mbox{\protect\boldmath $\alpha$}}

\newcommand{\bx}{\mbox{\protect\boldmath $x$}}
\newcommand{\by}{\mbox{\protect\boldmath $y$}}

\def\AmS{{\the\textfont2 A}\kern-.1667em\lower.5ex\hbox
     {\the\textfont2 M}\kern-.125em{\the\textfont2 S}}

\def\AW{Addison\kern.1em-\penalty0\hskip0pt Wesley}
\def\BibTeX{{\rm B\kern-.05em{\smc i\kern-.025emb}\kern-.08em\TeX}}
\author{Jakub Martinka}
\affiliation{J. Heyrovsk\'{y} Institute of Physical Chemistry, Academy of Sciences of the Czech \mbox{Republic, v.v.i.}, Dolej\v{s}kova 3, 18223 Prague 8, Czech Republic}
\alsoaffiliation{Department of Physical and Macromolecular Chemistry, Faculty of Science,
Charles University, Hlavova 8, 128 43 Prague 2, Czech Republic}

\author{Lina Zhang}
\affiliation[Second University]
{State Key Laboratory of Physical Chemistry of Solid Surfaces, Department of Chemistry, College of Chemistry and Chemical Engineering, and Fujian Provincial Key Laboratory of Theoretical and Computational Chemistry, Xiamen University, Xiamen 361005, China}

\author{Yi-Fan Hou}
\affiliation[Second University]
{State Key Laboratory of Physical Chemistry of Solid Surfaces, Department of Chemistry, College of Chemistry and Chemical Engineering, and Fujian Provincial Key Laboratory of Theoretical and Computational Chemistry, Xiamen University, Xiamen 361005, China}

\author{Mikołaj Martyka}
\affiliation[First University]
{Faculty of Chemistry, University of Warsaw, Pasteura 1, Warsaw, 02-093, Poland}

\author{Ji\v{r}\'{i} Pittner}
\email{jiri.pittner@jh-inst.cas.cz}
\affiliation{J. Heyrovsk\'{y} Institute of Physical Chemistry, Academy of Sciences of the Czech \mbox{Republic, v.v.i.}, Dolej\v{s}kova 3, 18223 Prague 8, Czech Republic}

\author{Mario Barbatti}
\email{mario.barbatti@univ-amu.fr}
\affiliation{Aix Marseille University, CNRS, ICR, Marseille, France}
\alsoaffiliation{Institut Universitaire de France, 75231 Paris, France}

\author{Pavlo O. Dral}
\email{dral@xmu.edu.cn}
\affiliation[Second University]
{State Key Laboratory of Physical Chemistry of Solid Surfaces, Department of Chemistry, College of Chemistry and Chemical Engineering, and Fujian Provincial Key Laboratory of Theoretical and Computational Chemistry, Xiamen University, Xiamen 361005, China}
\alsoaffiliation[Third University]
{Institute of Physics, Faculty of Physics, Astronomy, and Informatics, Nicolaus Copernicus University in Toru\'n, ul. Grudziadzka 5, 87-100 Toru\'n, Poland}
\alsoaffiliation[Aitomistic]
{Aitomistic, Shenzhen 518000, China}

\title[An \textsf{achemso} demo]
  {A Descriptor Is All You Need: Accurate Machine Learning of Nonadiabatic Coupling Vectors}

\abbreviations{IR,NMR,UV}

\begin{document}







\newpage
\begin{abstract}
Nonadiabatic couplings (NACs) play a crucial role in modeling photochemical and photophysical processes with methods such as the widely used fewest-switches surface hopping (FSSH). There is therefore a strong incentive to machine learn NACs for accelerating simulations. However, this is challenging due to NACs’ vectorial, double-valued character and the singularity near a conical intersection seam.
For the first time, we design NAC-specific descriptors based on our domain expertise and show that they allow learning NACs with never-before-reported accuracy of $R^2$ exceeding 0.99. The key to success is also our new ML phase-correction procedure.
We demonstrate the efficiency and robustness of our approach on a prototypical example of fully ML-driven FSSH simulations of fulvene targeting the SA-2-CASSCF(6,6) electronic structure level. This ML-FSSH dynamics leads to an accurate description of $S_1$ decay while reducing error bars by allowing the execution of a large ensemble of trajectories. Our implementations are available in open-source MLatom.

\end{abstract}

\section{Introduction}
It has been almost 100 years since quantum mechanics enabled a precise theoretical description of molecular systems, and early after its development, problems related to degenerate eigenvalues of the Hamiltonian were addressed\cite{Neumann2000}.
This became a crucial ingredient in simulating photophysical and photochemical processes involving nonradiative transitions from higher excited states to lower states via conical intersections (CI) or, in the case of different spin multiplicity, via intersystem crossings (ISC).
Unfortunately, even nowadays with the immense computing power and optimized algorithms, one is limited by the exponential scaling of fully quantum treatment. Hence, approximations leading to mixed quantum-classical approaches have to be introduced. One of these approximations, which is the focus of this work, is Tully's fewest-switches surface hopping (FSSH)\cite{Tully1990,HammesSchiffer1994}, which is a standard method for nonadiabatic dynamics simulations. It is based on separating classical and quantum degrees of freedom, integrating the quantum degrees with a locally approximated \mbox{time-dependent} Schr\"odinger equation, and allowing stochastic jumps between quantum states. 

FSSH requires a large amount, potentially hundreds of thousands, of single-point calculations of excited states, i.e., solving \mbox{time-independent} electronic Schr\"{o}dinger equation, limiting the dimensions (molecular size, number of trajectories, propagation time) of systems under inspection.
A desired solution is employing machine learning (ML) to conduct the simulations by fitting relevant quantities, circumventing the \mbox{time-consuming} quantum chemical calculations.
This effort started for ground-state properties many years ago \cite{Lorenz2004,Raff2005} and is a well-recognized workflow\cite{Dral2020,Behler2016,Deringer2019,Behler2021}, nevertheless, excited states pose additional challenges and the significant advances in simulating nonadiabatic processes were made only in the recent years\cite{Dral2018, Dral2021, Westermayr2019, Li2024a,Akimov2021,Atalar2024,Bai2022,Box2020,Boyer2024,Chen2018,Chen2019,Chen2020,Ghalami2024,Gherib2024,Holtkamp2023,How2021,Hu2023,Hu2018,Kraemer2020,Liu2024,Mazzeo2024,Posenitskiy2021,Richings2020,Tang2022,Ueno2021,Wang2025,Wang2023,Zhang2025a,Wu2025,Zhang2022}.

Within the FSSH framework, electronic energies, gradients, and nonadiabatic couplings (NACs) are necessary for propagating nonadiabatic molecular dynamics (NAMD) trajectories.
The potential energy surfaces (PESs) and their corresponding gradients are typically fitted together within a single ML model, whereas NACs require special treatment and a separate model.
The first-order nonadiabatic couplings ${\bm h}_{ij}$ are defined as
\begin{equation}\label{eq:NACs}
{\bm h}_{ij} = \bra{\Psi_i}\nabla_{\bm R}\ket{\Psi_j} = \frac{\bra{\Psi_i}\nabla_{\bm R}\hat{H}_{\text{el}}\ket{\Psi_j}}{E_i-E_j} \quad\text{for all} \quad i\neq j,
\end{equation}
where $E_i$, $E_j$ are the eigenvalues of electronic Hamiltonian $\hat{H}_{\text{el}}$ and $\Psi_i$ and $\Psi_j$ are the corresponding eigenstates. 
{NACs are used in three instances in FSSH. First, for the hopping probability, where they are employed to evaluate time-derivative couplings $\sigma_{ij}=\bra{\Psi_i}\frac{d}{dt}\ket{\Psi_j}=\bm{v} \cdot \bm{h}_{ij}$ \cite{pittner2009}, where $\bm{v}$ is the classical nuclear velocity. Second, they are used to evaluate whether hopping is allowed or frustrated. Third, they are required to rescale the velocity after hopping occurs \cite{Toldo2024,Plasser2019,Carof2017,Barbatti2021}.}

{NACs pose a significant challenge for FSSH, whether driven by quantum mechanics or machine learning \cite{Dral2018, Westermayr2020a, Hu2018}. Beyond their daunting vectorial nature (they constitute $N_s \left(N_s-1\right)/2$ vectors of $3N_a$ dimensions each for a molecular system with $N_s$ electronic states and $N_a$ atoms), NACs may exhibit discontinuities during dynamics due to arbitrary global phases of the electronic wave functions. They also tend to be narrow, Lorentzian-shaped functions near avoided crossings \cite{Baeck2017}, making them difficult to resolve in numerical integrations. This issue becomes severe in large systems, where near-delta-shaped NACs may emerge from couplings between spatially separated electronic states \cite{Wang2014a}. At CIs, NACs are formally divergent. A further complication arises from the geometric (Berry) phase \cite{Berry1984}: when the nuclear trajectory encircles a CI, NACs acquire a topological phase. If this phase is not tracked consistently, artificial discontinuities in the couplings are induced \cite{Ryabinkin2017}.}

{The simplest way to circumvent these complications is to avoid using NACs altogether. Alternative strategies include direct computation of the transition probabilities $\sigma_{ij}$~\cite{pittner2009} or local diabatization methods~\cite{granucci2001}, though both require explicit access to wave functions or their overlaps, which poses additional challenges for machine learning.}

{Approximated NAMD approaches offer another route, reducing the task to fitting only energies and gradients. They include methods such as Zhu--Nakamura~\cite{Ishida2017}, Landau--Zener~\cite{Zener1930, Landau1932, Landau1932a}, and time-dependent Baeck--An~\cite{Baeck2017, Casal2022} (also known as curvature-driven NACs~\cite{Shu2022}), all of which have been successfully applied in ML-driven simulations~\cite{Li2021, Li2022a, Li2022, Martyka2025, Sit2024, Li2025}. Although approximated NAMD schemes generally have a smaller impact on the outcome of the dynamics than the choice of the electronic structure method~\cite{Ibele2020, Janos2023}, they can still lead to noticeable deviations in excited-state behavior compared to FSSH trajectories using NACs. We will illustrate such a divergence between Landau--Zener and NAC-based surface hopping later, in the case of fulvene. These deviations arise from several factors: first, all these approximations rely on specific assumptions about the topology of the crossing seam when deriving hopping probabilities; second, evaluating frustrated hoppings without explicit NACs may lead to size-consistency issues~\cite{Toldo2024}; and third, rescaling velocities after hopping without knowing the NAC direction can result in applying nonadiabatic forces in the wrong direction~\cite{tully1998}.}

{Another option is to work in the diabatic representation \cite{Shu2024}, where, by definition NACs are null. In that representation, no divergences are observed in the Hamiltonian elements. However, it introduces the nontrivial task of identifying an appropriate diabatic basis. Machine learning can also assist here by learning diabatic representations directly from \textit{ab initio} data~\cite{Guan2017, Guan2019, Guan2020, Guan2020a, Shu2020, Axelrod2022, Li2023a, Srsen2024, Moon2025, Blasiak2022}.}

{For all these reasons, we believe that NAC-based FSSH remains the most reliable option for accurate surface hopping simulations, which motivates our development of machine learning models for NAC prediction.}

Focusing particularly on ML of NACs, Westermayr et al. obtained NACs as a first-order derivative of a virtual property by using an invariant SchNet\cite{Schuett2018} NN relying on structure-based\cite{Westermayr2019} or automatically designed descriptors\cite{Westermayr2020b} and later utilized an equivariant PaiNN\cite{Schuett2023} to predict the NACs directly\cite{Mausenberger2024a}.
Li et al. followed a similar direction and used NNs to fit atom-wise virtual potential and inverse distances, bond and dihedral angles as a descriptor\cite{Li2021a}.
Those studies focused on fitting the interstate coupling, the numerator in Eq.~\ref{eq:NACs}, removing the problem of singularities.

     
One of the most fundamental problems with ML of nonadiabatic couplings is the ambiguity in their sign caused by arbitrary phases of the bra and ket wave functions, causing an inconsistency in the training data set.
Within a single trajectory, the NACs phase can be tracked by extrapolation from previous steps, which was done with the help of wave function overlaps to the training set as well\cite{Westermayr2019}.
Another solution Westermayer et al. proposed is using the phaseless loss, a standard $L_2$ loss. Nevertheless, the loss is calculated $2^{N_S-1}$ times for $N_S$ states, which could be unbearable for an extensive system with many states\cite{Westermayr2020b}.
The approach recently proposed by Richardson\cite{Richardson2023} suggests learning the elements of the dyadic product of NACs.
This method was implemented in the DeepMD package\cite{Dupuy2024} but requires fitting of a rapidly growing matrix of $3N\times 3N$ elements.


The problem with all the above approaches is that the quality of the NAC vectors' fits with ML is relatively low, particularly compared with the quality of learning energies. We argue that the underlying reason is that
all previous studies on learning NACs have only used the standard molecular descriptors and ML techniques designed to learn energies, forces, and, sometimes, dipole moments while there is no guarantee that such choices are optimal for the specific problem of learning couplings.
The appropriate selection of features and ML models for each problem is very well known as an essential task for any ML undertaking, often requiring expert domain knowledge.

\begin{figure}
    \centering
    \includegraphics[width=.9\textwidth]{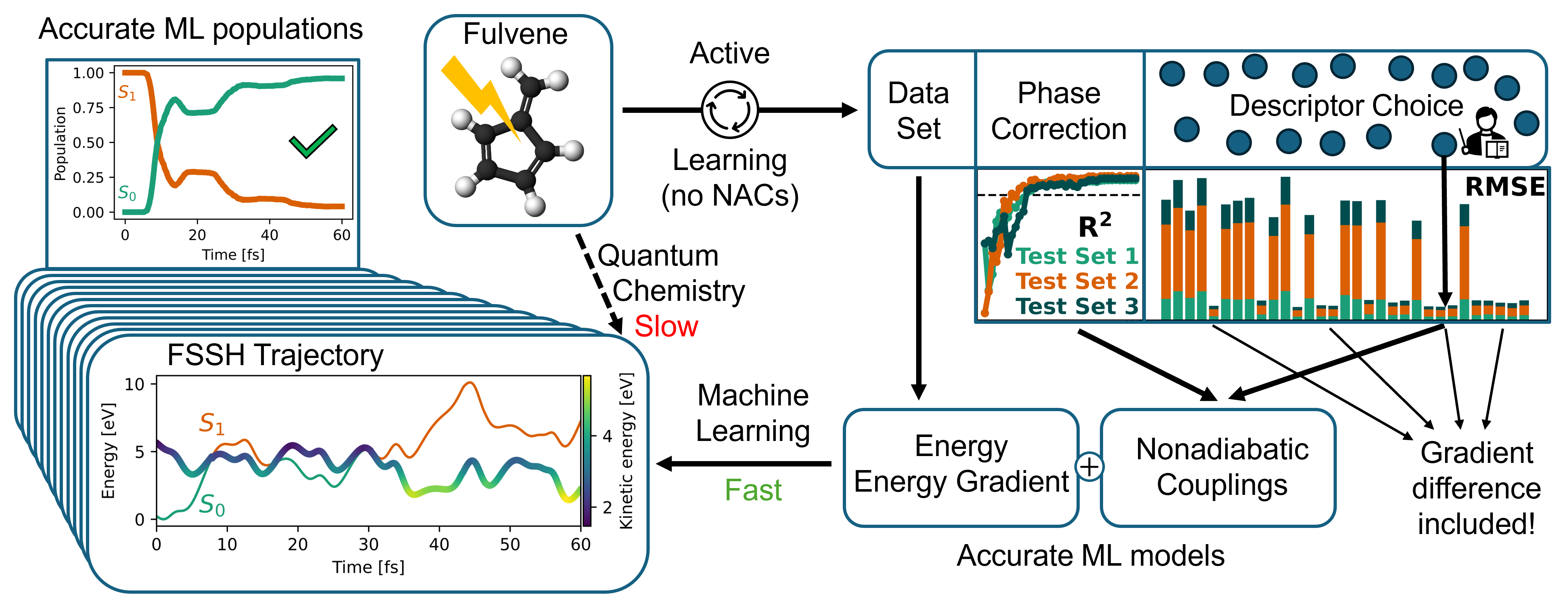}
    \caption{Schematic workflow of our ML-driven FSSH approach. We take the data set from Ref.~\citenum{Martyka2025}, which was generated through active learning with an MS-ANI model for energies and energy gradients employed in propagating Landau--Zener surface hopping dynamics and gapMD. Here, we apply phase correction to the data set to learn NACs, which are learned using kernel ridge regression. We show that incorporating gradient differences into the descriptor is essential for achieving high accuracy of ML NACs even without explicitly considering them within active learning. The ML models enable fast and precise predictions of all quantities required for FSSH, allowing for the execution of large ensembles of trajectories and reducing the statistical error in the final excited state populations.}
    \label{fig:chart}
\end{figure}

Here we approach the need for accurate learning of NACs by systematically analyzing the importance of descriptors crafted explicitly for this task (Fig.~\ref{fig:chart}). We also introduce a method for phase-correcting NAC data sets with ML, yielding the ML models for NACs with sufficient quality for high-precision FSSH dynamics. This methodological development extends the previously introduced active learning protocol for adequately sampling the molecular configurational space to build the training set for accurate 'NAC-free' LZSH dynamics performed with the ML models of MS-ANI type for predicting electronic energies and gradients.\cite{Martyka2025}
Ultimately, the combination of the high-accuracy ML models, ML-NACs developed here and MS-ANI developed previously, enables performing fully ML-driven FSSH. We test our approaches on an example of fulvene, which is a well-studied system\cite{Toldo2024,Jenkins2014,RuizBarragan2013,Sicilia2007,Tong2007,Blancafort2011,Blancafort2014,Casal2022,MendiveTapia2010,MendiveTapia2012}, considered as a molecular version of Tully's model III\cite{Ibele2020} and subjected to ML studies~\cite{Martyka2025,OMNIP2x}.


\section{Methods}
\subsection{Machine learning model for NACs}
In this work, we developed an iterative procedure for the phase-correction of NACs using kernel ridge regression (KRR)\cite{Hastie2009,Rupp2012} as an ML method capable of achieving high accuracy for relatively small data sets, while maintaining the training efficiency. The components of the NAC vectors are predicted as implemented in MLatom~\cite{MLatom2019,MLatom2024}:
\begin{equation}
    y(\bx') = \sum_{i=1}^{N_\text{tr}}\alpha_iK(\bx',\bx_i),
\end{equation}
where $N_{\text{tr}}$ is the number of training points, $\alpha_i$ is a regression coefficient, and $K$ represents the kernel function, which, in this study, has the Gaussian form
\begin{equation}\label{eq:Gauss}
    K({\bx}_i, {\bx}_j)=\exp\left(\frac{-||{\bx}_i-{\bx}_j||^2}{2\sigma^2}\right).
\end{equation}
In the training process, the regression coefficients are obtained by analytically solving the system of equations in matrix form $(\bK+\lambda I)\balpha=\by$, where $\bK$ is the kernel matrix calculated for all $N_\text{tr} \times N_\text{tr}$ pairs of the training points.
We also optimize the hyperparameters $\lambda$ and $\sigma$ for each training by using 20~\% of the training set as the validation set.
The optimization of hyperparameters was performed for all components of NACs together, i.e., $\by$ is a $N_\text{tr}\times 3N_\text{atoms}$ matrix, and the solution provides the matrix of the regression coefficients $\balpha$ of the same size and shape. This approach has the advantage of lowering the cost of training by ca.~$3N_\text{atoms}$ times because the most time-consuming part is the Cholesky decomposition of the kernel matrix, which is built once for all NACs components. The disadvantage is that we use the same hyperparameters for each NAC component, which may lower the accuracy of the fitting. A similar approach of learning all couplings at once was reported earlier in a different context of learning several quantum chemical properties\cite{Ramakrishnan2015}.

The accuracy of ML models strongly depends on the input features $\bx$, whose choice is the focus of this work and will be discussed later.

\subsection{Phase-correction with machine learning}
\begin{figure}
    \centering
    \includegraphics[width=.7\textwidth]{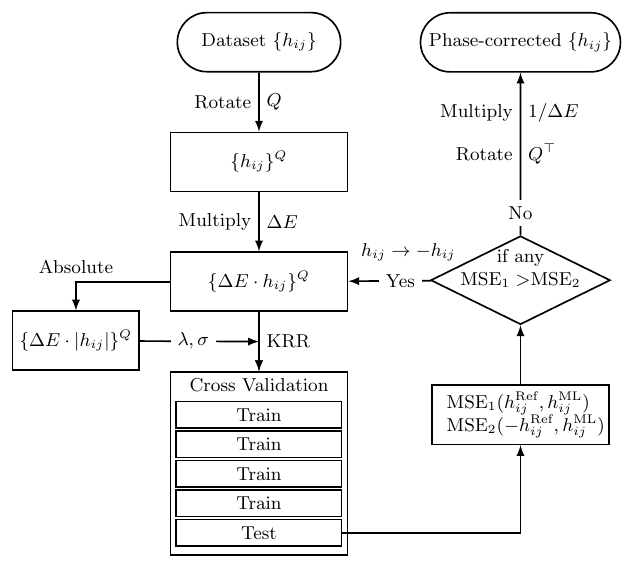}
    \caption{Phase-correcting procedure of NAC vectors introduced in this work. In the first step, couplings are rotated into a selected reference frame and then scaled by the energy gap. An iterative process is employed where a KRR model is retrained in each iteration. Initially, absolute NAC values are used to determine hyperparameters fixed within the scheme to enhance the procedure. The training uses a 5-fold cross-validation method, with NACs predicted for the remaining part of the data set. Mean squared error is calculated between predictions and positive or negative reference values. If the MSE corresponding to negative reference values is lower than that of the positive ones, the sign of the corresponding NAC in the training set is changed. The iterative procedure continues until there are no sign flips or the process has converged within a predefined patience threshold.}
    \label{fig:diagram}
\end{figure}

In this Section, we describe our procedure for phase-correcting NAC vectors to enable their accurate learning; the procedure is illustrated in a diagram in Fig.~\ref{fig:diagram}.
We utilize the rotation to the local-frame orientation and the KRR models, as they are the most efficient approaches available for our learning objectives from as small data sets as possible\cite{Pinheiro2021}.
In the first step, the rotational matrices ($Q, Q^\top$) are obtained via the Kabsch algorithm\cite{Kabsch1976}, where equilibrium geometry is used as a reference and the properties of vectorial nature (gradients and NACs) are rotated for all the molecules in the data set, following a Rotate–Predict–Rotate approach\cite{Martinka2024}.
Next, NACs are multiplied by the energy gap to eliminate singularities, as was previously reported\cite{Westermayr2020b}, which corresponds to fitting the numerator in Eq.~\ref{eq:NACs}.
We aim to phase-correct the set using an iterative procedure, where the KRR model is retrained in each iteration.
To enhance efficiency, we first train the KRR model on the absolute values of NACs to get the hyperparameters $\lambda, \sigma$, and fix them for other iterations.
The KRR models for correction are trained in a 5-fold cross validation, where randomly selected 80~\% of the data is used for training, and the remaining 20~\% of the NACs are predicted.
Each NAC prediction (${\bm h}_{ij}^{\text{ML}}$) is then compared with the positive and negative reference values (${\bm h}_{ij}^{\text{Ref}}$) by calculating the mean square error (MSE)
\begin{equation}
    \text{MSE}\left({\bm h}_{ij}^{\text{Ref}}, {\bm h}_{ij}^{\text{ML}}\right) = \frac{1}{3N}\sum_{A=1}^{3N}\left({\bm h}_{ij}^{\text{Ref}}(A)-{\bm h}_{ij}^{\text{ML}}(A)\right)^2
\end{equation}
for a system of $N$ atoms.
If the condition
\begin{equation}\label{eq:MSE}
    \text{MSE}\left({\bm h}_{ij}^{\text{Ref}}, {\bm h}_{ij}^{\text{ML}}\right) > \text{MSE}\left(-{\bm h}_{ij}^{\text{Ref}}, {\bm h}_{ij}^{\text{ML}}\right)
\end{equation}
is satisfied, ${\bm h}_{ij}^{\text{Ref}}$ is multiplied by $-1$ and the process is repeated for each fold of the \mbox{cross-validation}.
If any sign flip occurs across the 5 folds, the procedure continues until there are no sign flips or the process has converged within a predefined patience threshold. 


\subsection{ML-driven fewest-switches surface hopping}
The FSSH approach requires energy, energy gradients, and NACs.
For calculating energies and energy gradients with ML, we use the MS-ANI model successfully employed for fulvene earlier~\cite{Martyka2025}. A KRR model predicts the NACs.
During FSSH simulations, MS-ANI predicts energies and energy gradients for a given molecular geometry ${\bR}(t)$.
If the descriptor includes vectorial properties, such as gradient difference (see below), they are rotated consistently with the phase-correction procedure using the rotational matrix $Q$ obtained via the Kabsch algorithm.
The KRR model then predicts ${\bh}_{ij}^{\text{ML}}$ for ${\bR}(t)$, which is subsequently rotated back with $Q^\top$ and divided by $\Delta E^{\text{ML}}_{ij}$.
To ensure phase consistency between consecutive classical time steps $t$ and $t+\Delta t$, the sign of ${\bh}_{ij}^{\text{ML}}(t+\Delta t)$ is changed in the case of negative value of scalar product between ${\bh}_{ij}^{\text{ML}}(t)$ and ${\bh}_{ij}^{\text{ML}}(t+\Delta t)$.
If hopping occurs during the propagation, the velocity is rescaled in the direction of NACs.

In our experiments described below, the initial conditions are sampled from a harmonic oscillator Wigner distribution, and the dynamics started from the $S_1$ state.

\section{Results and discussion}
We tested our ML models on the prototypical fulvene molecule.
From the photochemical point of view, fulvene is a representative for double-bond isomerisation, but it is also challenging due to its extended crossing seam with both sloped and peaked CIs causing ultrafast population transfer from $S_1$ to $S_0$ with a following partial transfer back to $S_1$\cite{Casal2022}.
The crossing seam corresponds to the rotation of the methylene group with saddle point at planar/perpendicular orientation with the minimum energy CI (MECI) located at the angle of 63°\cite{Blancafort2014}.
Nevertheless, the decay occurs mainly at the planar orientation\cite{Blancafort2011}.

To evaluate ML models' quality and performance within FSSH, we use similar settings as in Ref.~\citenum{Toldo2024}.
All NAMD simulations have been performed using the newly implemented FSSH within the MLatom software~\cite{MLatom2024,Zhang2024}, which runs natively in Python, overcoming the inefficiency of read/write operations, increasing the convenience of using ML models, and enabling easy data analysis of the resulting trajectories in a memory-efficient binary H5MD format \cite{Buyl2014}.
To generate the reference data (energies, energy gradients, and NAC vectors) for ML training, we employed the SA-2-CASSCF(6,6) with the 6-31G(d) basis set using MLatom's interface to Columbus (version 7.2, 2022)\cite{Columbus7.2,Lischka2020}. The same method was used to generate the reference FSSH trajectories for comparison.

In the following, we describe the choice of descriptors for the KRR model, convergence of the algorithm for phase-correction, quality of NAC predictions, and, finally, fully ML-driven NAMD simulations. The Velocity Verlet algorithm with a time step of 0.1 fs was used to propagate classical degrees of freedom up to 60 fs.
Time-dependent Schr\"{o}dinger equation was integrated with 4th order Runge-Kutta algorithm with a time step of 0.005 fs. To account for the decoherence, the simplified decay of mixing (SDM)\cite{Granucci2007} correction with $\alpha$ parameter set to 0.1~au was applied.

\subsection{Descriptor choice and the quality of the machine learned nonadiabatic couplings}\label{subsec:descriptor}

The accuracy of ML models strongly depends on the input features $\bx$, and their selection is inherently task-dependent. In quantum chemistry, molecular structures must be converted into features called molecular descriptors. Nevertheless, all previous studies only considered descriptors developed for machine learning energies and energy gradients. The latter features are derived from the 3D molecular geometries, and such structural descriptors can conveniently be divided into local and global; they are typically constructed to achieve rotational, translational, and, if possible, permutational invariance or equivariance.

Here, we address the gap in identifying suitable descriptors for NAC vector learning by using our domain expertise to guide the selection of promising candidates.
While there is a plethora of possible structural descriptors, we limit ourselves to the relative-to-equilibrium (RE)\cite{Dral2017} representation, which proven itself one of the best choices for constructing accurate KRR-based KREG models~\cite{Hou2023} for energies and energy gradients, capable of achieving spectroscopic accuracy~\cite{Dral2017,Dral2020a}. Among domain-specific descriptors potentially important for learning NAC vectors, we consider energy difference ($\Delta E$) and gradient difference ($\Delta\nabla E$) between the corresponding pair of states.
The choice of $\Delta E$ is motivated by the difference between energy levels influencing the magnitude of NACs.
Regarding $\Delta\nabla E$, the difference between energy gradients together with the NAC form a 2-dimensional branching space around the CI, while two energy levels are degenerate within the remaining ($3N-8$)-dimensional subspace of nuclear coordinates\cite{ConicalIntersections2004}.
In particular, noting that at the crossing seam the two states become exactly degenerate, one has the freedom of unitary transformations, which keep the states degenerate. 
Such transformations mix the vectors of gradient difference between the two states and their NAC vector (however, keeping the plane defined by these two vectors invariant)\cite{Meisner2015}.
The gradient differences computed at many points near the CI seam, which are part of the training set, should thus sample the whole plane defined by the gradient difference and NAC vectors at the exact CI seam. Hence, gradient differences form a potentially useful descriptor for ML of the NAC vectors. The $\Delta\nabla E$ descriptor is not rotationally invariant. Therefore, they are rotated using the rotational matrix $Q$ obtained from the Kabsch algorithm with respect to the equilibrium geometry.


In addition to the RE, energy difference ($\Delta E$), and gradient difference ($\Delta\nabla E$) descriptors, we also consider 
absolute value of gradient difference ($|\Delta\nabla E|$) and Frobenius norm  of a gradient difference ($||\Delta\nabla E||_F$) defined for a $m\times n$ matrix $\bA$ as
\begin{equation}
    ||{\bm A}||_F = \sqrt{\sum_{i=1}^m\sum_{j=1}^n|{\bm A}_{ij}|^2}.
\end{equation}


To benchmark the importance of the above descriptors, we assemble three different test sets from 200 CASSCF trajectories of fulvene.
The first set consists of 5000 randomly selected points from all trajectories, the second set contains 5000 randomly selected points, where the Frobenius norm of NACs is larger than 5, and the third one is created by taking all the data of the first 10 trajectories, i.e., a total of 6010 points. 
The quality of descriptors and all their possible combinations as input features is evaluated on these test sets.
We particularly emphasize the importance of the second test set, consisting of points in the configurational space with a large magnitude of the NACs, a critical region where nonadiabatic transitions occur, which must be accurately described.
We trained the KRR models using the above input features on the fulvene data set with 5950 points obtained via active learning from Ref.~\citenum{Martyka2025}.

\begin{figure}
    \centering
    \includegraphics[width=.9\textwidth]{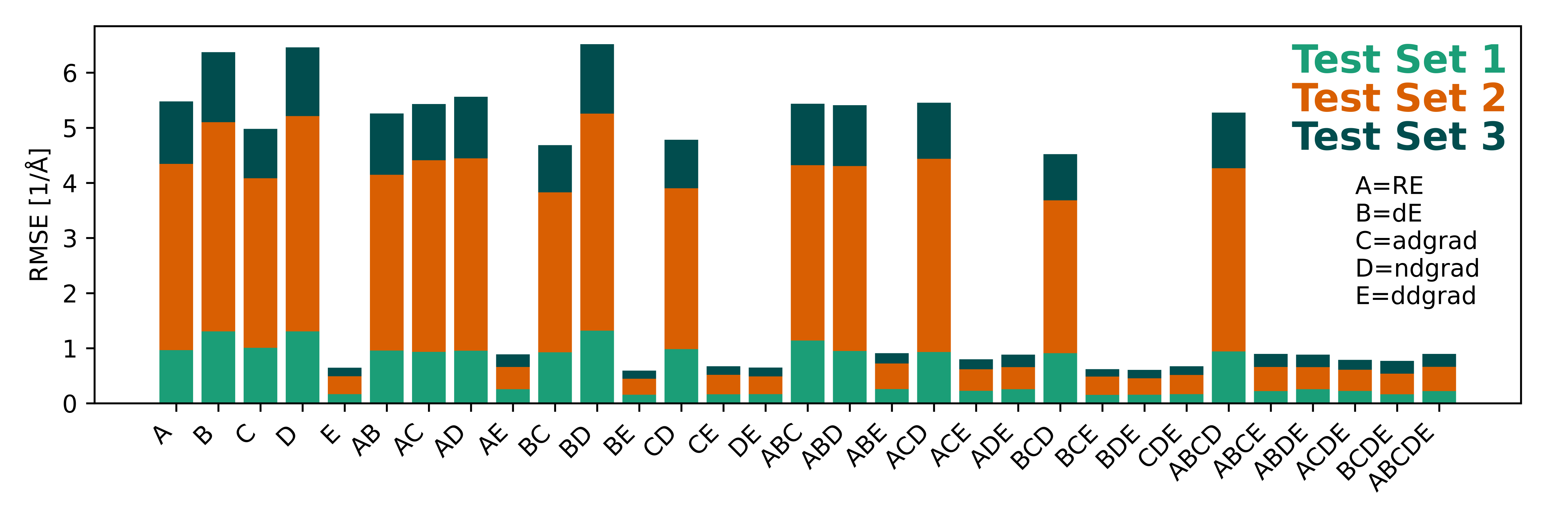}
    \caption{Descriptor benchmark showing RMSE values for all combinations of investigated descriptors. Relative to equilibrium (RE), energy difference (dE, $\Delta E$), gradient difference (ddgrad, $\Delta\nabla E$), absolute value of gradient difference (adgrad, $|\Delta\nabla E|$), and Frobenius norm of gradient difference (ndgrad, $||\Delta\nabla E||_F$). The error evaluated on three test sets is significantly decreased if ddgrad descriptor is included.}
    \label{fig:descriptors}
\end{figure}

The comparison in Fig.~\ref{fig:descriptors} clearly shows that the $\Delta\nabla E$ is by far the most important descriptor
(see Supporting Information for a complete comparison). The root-mean-squared error (RMSE) averaged over all test sets is 0.215~1/{\AA} and the coefficient of determination ($R^2$) is close to 0.99. Including other descriptors only marginally influences the $R^2$; the best performing combination is $\Delta\nabla E$ with $\Delta E$, which gives $R^2$ slightly higher than 0.99.
This is a remarkable result that proves that NACs can be accurately learned.

Based on the principle of parsimony, our final recommendation for the descriptors for \mbox{ML-NAC} models is the gradient difference ($\Delta\nabla E$), which showed the highest importance among all tested features.
This descriptor, as we will show below, is well-suited for the ML-FSSH dynamics.
In some cases, such as phase-correction, including more descriptors might improve results, i.e., the RE descriptor together with gradient difference performed the best for smaller subsets used for preliminary testing and resulted to robust and fast convergence of the procedure. Even in this case, the key ingredient is the presence of gradient difference and the RE descriptor can be safely dropped for simplicity.

\subsection{Convergence of phase correction}
Here, we employ the KRR model with the RE combined with CASSCF $\Delta\nabla E$ descriptor for the phase correction procedure. We test the performance of this procedure
using the properties calculated with CASSCF for the aforementioned 5950-point fulvene data set obtained from active learning\cite{Martyka2025}.

\begin{figure}
    \centering
    \includegraphics[width=.9\textwidth]{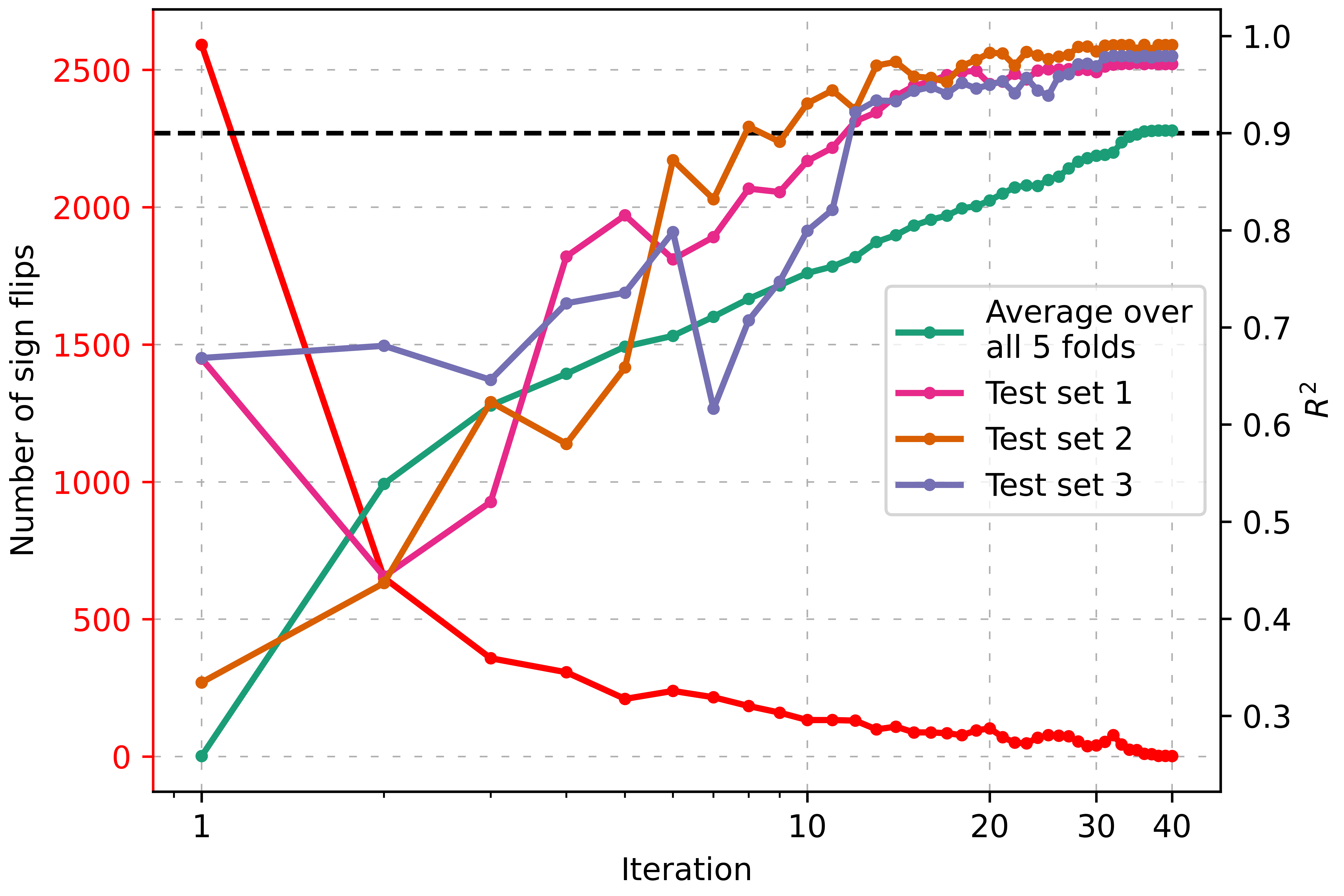}
    \caption{Phase correction of a 5950-point data set using KRR model with RE and $\Delta\nabla E$ descriptor. Phase correction has been performed iteratively with 5-fold cross-validation.
    In each iteration, KRR is trained with pre-defined hyperparameters.
    The number of sign flips (red) decreases steadily, while the average correlation across folds (light green line) increases, indicating improved phase consistency. KRR models retrained with hyperparameter optimization on intermediate data sets are evaluated on three test sets (magenta, orange, and purple lines) and surpass a correlation of 0.9 within 12 iterations and converge above 0.99 in the case of the crucial test set 2, demonstrating the effectiveness of the phase correction procedure.}
    \label{fig:phcorr}
\end{figure}

Each iteration of our phase-correcting procedure (Fig.~\ref{fig:diagram}) invokes a 5-fold cross-validation (CV) approach, where the KRR model, with predefined hyperparameters $\lambda$ and $\sigma$.
As shown in Figure~\ref{fig:phcorr}, the number of sign flips progressively decreases (red), while the average $R^2$ across all folds (light green line) consistently increases throughout the phase-correction procedure. For this data set, the process has converged in 40 iterations.

For further improvement, we save the phase-corrected data sets from intermediate iterations and retrain the KRR models this time by performing hyperparameter optimizations.
We assess the accuracy of these models on three test sets described above in the context of choosing descriptors.
As seen in Figure~\ref{fig:phcorr}, the KRR models surpass $R^2$ of 0.9 within just 12 iterations and ultimately converge to $R^2$ of 0.99 in the case of the crucial test set 2.
This achievement highlights the critical role of precise phase correction in enhancing ML model accuracy, demonstrating the effectiveness of our approach.

\subsection{Performance of machine learned NACs in FSSH}
The real test of the ML models comes from their performance in the actual simulations of interest, i.e., it is not enough to analyze in-distribution errors on static data sets~\cite{Ge2024}. In our case, such a test is the performance of ML-NACs in FSSH dynamics. As the FSSH quality metric, we will analyze the model's performance on the $S_1$ state populations, which reflect the number and timing of hoppings strongly influenced by the quality of the NACs predictions. However, there is a serious complication in such an analysis because ML-NACs alone are insufficient for performing the FSSH simulations, which also require calculation of energies and energy gradients needed to propagate trajectories and derive the descriptors (i.e., gradient difference $\Delta\nabla E$ is part of them) for ML-NACs. Our ultimate goal is a fully ML-driven FSSH, where ML is used for predicting NACs, energies, and energy gradients, but it is crucial to analyze the quality of the ML model(s) for each of these properties.

\begin{figure}
    \centering
    \begin{overpic}[width=.9\textwidth]{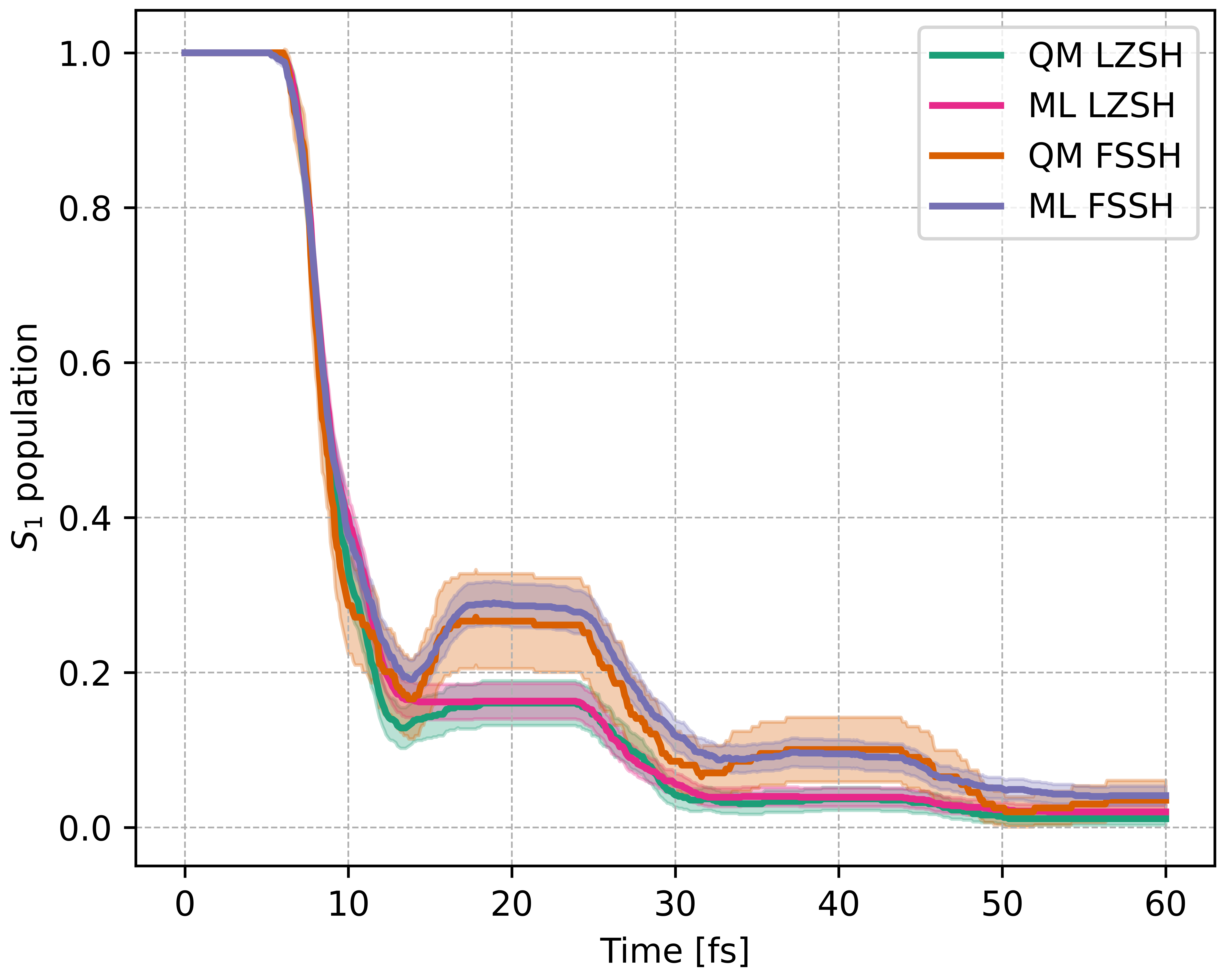}
        \put(70,30){\includegraphics[height=.15\textwidth]{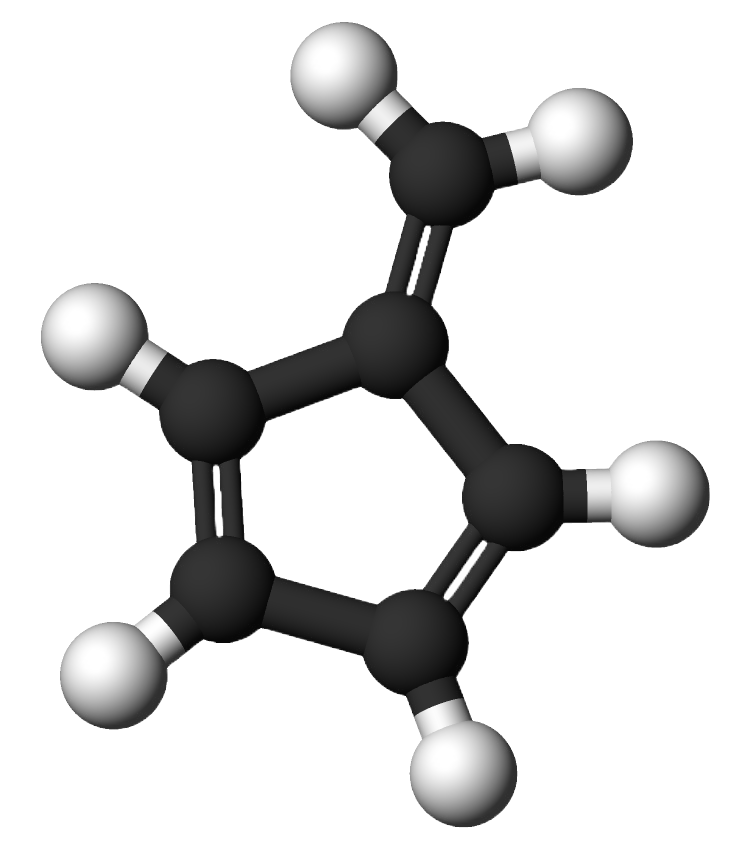}}
    \end{overpic}
    \caption{Excited-state population of fulvene computed using FSSH and LZSH schemes.
    (a) The hybrid scheme that combines CASSCF for energies and gradients with machine-learned NACs accurately reproduces the pure CASSCF population, demonstrating the quality of the ML fit. This scheme is denoted "QM($E$, $\nabla E$) + ML-NAC(ddgrad)". The population predicted by the MS-ANI and ML-NAC models (denoted "MS-ANI + ML-NAC(descriptor)") strongly depends on the choice of descriptor. When using the Relative to Equilibrium (RE) descriptor, the population is not correctly captured, whereas the $\Delta\nabla E$ (ddgrad) descriptor yields excellent agreement with the reference pure CASSCF population.
    (b) The importance of NACs in NAMD simulations is demonstrated by comparing the Fewest Switches Surface Hopping (FSSH) approach with the Landau--Zener Surface Hopping (LZSH) method, which requires only energies and gradients. A key advantage of ML-driven NAMD is the ability to simulate a large ensemble of trajectories (1000 ML vs. 200 QM), reducing the confidence intervals' width. LZSH population data were taken from Ref.~\citenum{Martyka2025}.
    Reference trajectories were computed at the SA-2-CASSCF(6,6)/6-31G(d) level of theory. In FSSH simulations, state coefficients were corrected for decoherence using the SDM method, and upon hopping, velocities were rescaled along the direction of the NACs.
    }
    \label{fig:pop}
\end{figure}

Hence, before we discuss a fully ML-driven FSSH, we first analyze whether the quality of the ML model for NACs is sufficient for the accurate FSSH dynamics. For this, we use the reference CASSCF method to calculate energies and energy gradients on-the-fly during the FSSH trajectory propagation to integrate the equations of motion and derive the $\Delta\nabla E$ descriptor for the ML-NAC model. We take the ML-NAC model obtained after the phase correction on the 5950-point fulvene data set with the CASSCF-level NACs (learning target) and energy gradients (used to derive the descriptor $\Delta\nabla E$).
This data set was generated through a newly implemented active learning framework incorporating uncertainty quantification and a sampling procedure using gapMD~\cite{Martyka2025}.
However, the data set has been collected based on the NAMD dynamics performed within the Landau--Zener approximation rather than with the FSSH algorithm. Encouragingly, the $S_1$ state populations with such a hybrid FSSH approach employing ML-NACs and CASSCF energies and energy gradients agree very well with the reference fully CASSCF populations (Fig.~\ref{fig:pop}a), considering the stochastic nature of FSSH and the relatively wide error bars due to the limited number of trajectories (Fig.~\ref{fig:pop}b). This experiment shows that the quality of the ML-NACs model is excellent, and this model can enable the robust FSSH dynamics.
This demonstrates that an active learning procedure based on an approximate and faster Landau--Zener can produce a data set suitable for training reliable ML-NACs models transferable to FSSH.

But how does such an ML-NAC model perform with the ML model for energies and energy gradients? We take the MS-ANI model from Ref.~\citenum{Martyka2025} trained on the same data set as the ML-NAC model to answer this question. We use this MS-ANI model for on-the-fly calculations of energies and energy gradients employed in propagating the FSSH trajectories and calculating the gradient difference $\Delta\nabla E$ descriptor for the ML-NAC model. 
The resulting $S_1$ state population (Fig.~\ref{fig:pop}a) is in excellent agreement with reference CASSCF when $\Delta\nabla E$ descriptor is included. In contrast, the model based on the structural descriptor does not capture the correct population dynamics.
{We observe in Fig.~\ref{fig:pop}b that LZSH have some noticeable deviations in excited-state decay, compared to the FSSH trajectories employing NACs. In particular, back hoppings repopulating $S_1$ around 12~fs is not correctly captured by the Landau--Zener approach, which otherwise performs well within the singlet manifold of states~\cite{Suchan2020}. This finding underscores the importance of having NAC-based dynamics available.} 

The greatest advantage of ML-driven trajectories is their high speed---in our case, they are 434 times faster than conventional CASSCF trajectories, with room still left for further optimization.
For the fully ML-driven NAMD, 1000 initial conditions can be used for the propagation as they are computationally inexpensive, while in the case of CASSCF, only the first 200 initial conditions have been used. 
As can be seen from Table~\ref{tab:mean}, using fully-ML driven NAMD simulations enables the calculation of a large ensemble of trajectories, which can significantly decrease the error bars of confidence interval~-- similarly to the previously reported KRR protocol for the UV/VIS spectra via nuclear ensemble approach\cite{Xue2020}.

\begin{figure}
    \centering
    \includegraphics[width=.9\textwidth]{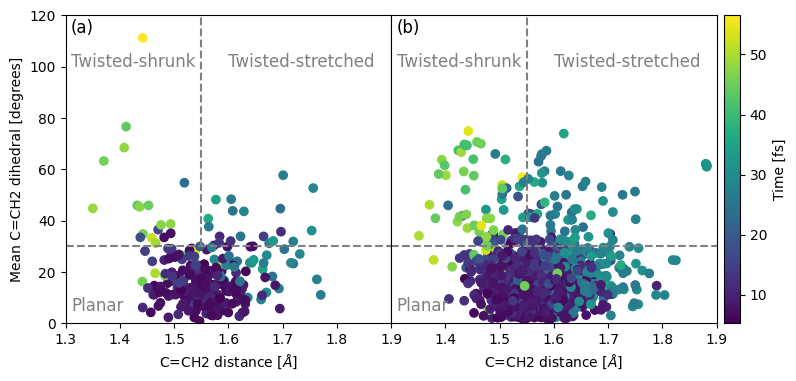}
    \caption{Correlation plots of C$=$CH$_2$ dihedral angle and distance characterizing CASSCF (a) and ML (b) dynamics of fulvene. The first transfer from $S_1$ to $S_0$ occurs between 7 and 14 fs at the planar structure (dark blue points). Next, reverse transfer back to $S_1$ follows, and hopping to $S_0$ occurs again between 20 and 40 fs at either planar or twisted-stretched structure (light blue points). The terminal stage of the simulation corresponds to hopping close to the MECI of 63° C$=$CH$_2$ dihedral angle at twisted-shrunk structure (light green points).}
    \label{fig:hoppgeoms}
\end{figure}

\begin{table}[htbp]
\caption{Mean value and error bars (95~\% confidence interval) of observables describing the deactivation channels and kinetics of fulvene for ML dynamics and reference CASSCF dynamics.}
\begin{tabular}{lcccc}\toprule
     & \multicolumn{2}{c}{LZSH} & \multicolumn{2}{c}{FSSH}\\
    Observable & {MS-ANI} & {CASSCF} & {MS-ANI + KRR} & {CASSCF}\\
    \midrule
    Population at 20 fs ($\%$) & 16.3 $\pm$ 2.3 & 16.1 $\pm$ 2.9 & 28.3 $\pm$ 2.8 & 26.6 $\pm$ 6.1  \\
    Population at 40 fs ($\%$) & 3.9 $\pm$ 1.2 & 3.7 $\pm$ 1.5 & 9.7 $\pm$ 1.8 & 10.1 $\pm$ 4.2 \\
    Planar hopping ($\%$) & 93.8 $\pm$ 1.5 & 94.0 $\pm$ 1.8 & 90.4 $\pm$ 1.6 & 85.2 $\pm$ 4.2 \\
    Twisted-stretched hopping ($\%$) & 4.1 $\pm$ 1.2 & 3.7 $\pm$ 1.8 & 5.5 $\pm$ 1.2 & 7.7 $\pm$ 3.2\\
    Twisted-shrunk hopping ($\%$) & 2.2 $\pm$ 0.9 & 2.3 $\pm$ 1.2 & 4.1 $\pm$ 1.1 & 7.0 $\pm$ 3.0\\\bottomrule
\end{tabular}
\label{tab:mean}
\end{table}

Moreover, initial conditions may play a crucial role in describing photochemical processes as well \cite{Janos2025}.
As shown in Table~\ref{tab:mean}, the percentage of hopping channels varies when a larger sample of initial conditions is considered.
Nevertheless, Fig.~\ref{fig:hoppgeoms} demonstrates that the fully ML-driven FSSH simulations correctly capture the hopping events at the planar structure, corresponding to the $S_1\rightarrow S_0$ transition occurring between 7 and 14 fs.
Subsequently, another hopping takes place between 20 and 40~fs, occurring at either a planar or twisted-stretched structure, and in the terminal stage, hopping occurs at the twisted-shrunk structure.

\section{Conclusions}
In this study, we proposed suitable descriptors for efficient and robust machine learning of nonadiabatic coupling (NAC) vectors. These descriptors were designed based on domain knowledge and, for the first time, include features beyond standard structural descriptors. We identified the gradient difference as the most relevant feature for learning NACs, enabling the construction of models with $R^2$ values exceeding 0.99. A crucial element of our approach is a new phase-correction procedure built upon these descriptors and implemented within a KRR-based framework.

We performed extensive tests of the ML-NAC model performance in the FSSH simulations of the prototypical fulvene molecule and discovered that special care has to be taken when choosing the descriptor, because the performance of ML-driven FSSH and the resulting $S_1$ decay strongly depends on the ML-NAC model. 

Interestingly, in our tests we have used the training sets and ML model for energies and gradients which were obtained via active learning targeting the 'NAC-free' LZSH dynamics, which is a faster but more approximate type of NAMD than FSSH. The high-quality of the resulting ML-FSSH dynamics indicates that the training sets and ML models obtained for one type of the NAMD might be transferable to other types. This is encouraging as we may reuse the data sets for different variants of NAMD and, e.g., use the faster variants to perform resource-intensive active learning. However, it is also possible to extend the previous active learning protocol developed for LZSH to FSSH with a few modifications incorporating the protocol for ML-NACs introduced here.


To summarize, the recipe for the robust fully ML-driven FSSH dynamics is accurate machine learning of NACs with the gradient difference descriptor, proper NAC phase correction for training, and gradients accurately predicted with ML for critical points near the conical intersections. {Our implementation, available in open-source MLatom, provides a general framework for accurate and efficient ML-driven nonadiabatic dynamics. It can be later extended to accelerate other variants of NAMD requiring NACs such as \textit{ab initio} multiple spawning\cite{AIMS2000}, mapping approach to surface hopping (MASH),\cite{Mannouch2023} and
nonadiabatic field (NAF)\cite{He2024,Wu2024,Wu2025}.}

\subsection{Data and code availability}
Fulvene data set generated in this work, as well as machine learning models (MS-ANI, KRR) are available from FigShare: \url{https://doi.org/10.6084/m9.figshare.28877672}.
Jupyter notebook used for the analysis, corresponding data, and any additional code are available from \url{https://github.com/JakubMartinka/Fulvene-ML-FSSH}.
The code for FSSH, enabling ML-FSSH, is available in open-source MLatom under the MIT license in \url{https://github.com/dralgroup/mlatom}.

\subsection{Authors contributions}
J.M. implemented the final version of FSSH, phase-correction algorithm, performed final calculations and their analysis, wrote the original manuscript.
L.Z. made original implementations and tests of phase-correction protocols.
Y.F.H. implemented the final version of the KRR method.
M.M. obtained the training set for fulvene dynamics, trained the MS-ANI model for PES', and contributed to the analysis and debugging of the FSSH code.
J.P. co-designed and co-conceived the project, contributed to the result analysis and interpretation, co-supervised research, and secured funding.
M.B. co-designed and co-conceived the project, co-designed the descriptor choice, contributed to the result analysis and interpretation, co-supervised research, and secured funding.
P.O.D. co-designed and co-conceived the project, co-designed descriptor choice and machine learning protocols, did intermediate implementations and tests of KRR and phase-correction algorithms, contributed to the result analysis and interpretation, co-supervised research, and secured funding.
All authors discussed the results and revised the manuscript.

\begin{acknowledgement}
The authors acknowledge Bao-Xin Xue for assisting with the ancient version of the phase-correction implementation.

The work of the Czech team has been supported by the \textit{Czech Science Foundation} Grant \mbox{23-06364S}, the \textit{Charles University} (project GAUK 6224) and
by the Advanced Multiscale Materials for Key Enabling Technologies project of the \textit{Ministry of Education, Youth, and Sports of the Czech Republic}, project No. CZ.02.01.01/00/22\_008/0004558, Co-funded by the European Union.
We also highly appreciate the generous admission to computing facilities owned by parties and projects contributing to the National Grid Infrastructure MetaCentrum provided under the program ``Projects of Large Infrastructure for Research, Development, and Innovations ''(no. LM2010005) and computer time provided by the IT4I supercomputing center (Project ID:90254) supported by the Ministry of Education, Youth and Sports.
M.M. acknowledges the Polish Ministry of Education and Science for funding this research under the program “Perły Nauki,” grant number PN/01/0064/2022, amount of funding and the total value of the project: 239 800,00 PLN, as well as gratefully acknowledge Polish high-performance computing infrastructure PLGrid (HPC Centers: ACK Cyfronet AGH) for providing computer facilities and support within computational grant no. PLG/2024/017363.
M.B. thanks the funding provided by the European Research Council (ERC) Advanced grant SubNano (Grant agreement 832237). M.B. received support from the French government under the France 2030 as part of the initiative d'Excellence d'Aix-Marseille Universit\'e, A*MIDEX (AMX-22-IN1-48).
P.O.D. acknowledges funding from the National Natural Science Foundation of China (via the Outstanding Youth Scholars (Overseas, 2021) project), via the Lab project of the State Key Laboratory of Physical Chemistry of Solid Surfaces, and Aitomistic, Shenzhen. Part of the computations were performed using the XACS cloud computing resources.

\end{acknowledgement}





\providecommand{\latin}[1]{#1}
\makeatletter
\providecommand{\doi}
  {\begingroup\let\do\@makeother\dospecials
  \catcode`\{=1 \catcode`\}=2 \doi@aux}
\providecommand{\doi@aux}[1]{\endgroup\texttt{#1}}
\makeatother
\providecommand*\mcitethebibliography{\thebibliography}
\csname @ifundefined\endcsname{endmcitethebibliography}
  {\let\endmcitethebibliography\endthebibliography}{}

\end{document}